\documentclass[12pt,a4paper]{article}
\usepackage{amsmath,amssymb,booktabs,array,placeins,graphicx,float}
\usepackage[numbers,sort&compress]{natbib}
\usepackage{url}
\usepackage[a4paper,textwidth=6.8in,top=25mm,bottom=25mm,centering]{geometry}
\begin{document}
\begin{center}
{\Large\bfseries Physics-informed neural network (PINN) approximations of capillary menisci in polygonal channels\par}
\vspace{0.8em}
Zijian Liu, Julian Olszewski, and Jie Xu\textsuperscript{*}\\[0.5em]
{\small Department of Mechanical and Industrial Engineering, University of Illinois Chicago,\\ 842 W. Taylor Street, Chicago, IL 60607, USA}
\end{center}
\begingroup
\renewcommand{\thefootnote}{}
\footnotetext[0]{Zijian Liu and Julian Olszewski contributed equally to this work.}
\endgroup
\begingroup
\renewcommand{\thefootnote}{\fnsymbol{footnote}}
\footnotetext[1]{Corresponding author: jiexu@uic.edu.}
\endgroup
\vspace{0.5em}
\noindent\textbf{Abstract}\par\smallskip
\noindent A capillary interface must satisfy both its curvature equation and the contact angle at the container walls. We study how strong-form physics-informed neural networks (PINNs) approximate these conditions when the prescribed equilibrium is geometrically incompatible. In zero gravity, the Concus--Finn critical contact angles for spanning graphs in square and equilateral-triangular cross-sections are respectively \(45^\circ\) and \(60^\circ\). Below these angles, classical subdomain flux balance yields a certified lower bound on the continuous combined interior and wall residual loss, retaining the sampling densities and wall weight. The subdomain bounds and their sums over disjoint regions tend uniformly to zero as the contact angle approaches the threshold from below at fixed wall weight. The computed PINNs return smooth graphs with finite axial extent in the nonexistence regime: small interior residuals coexist with corner wall-flux deficits, where the adjoining contact-angle conditions demand more flux than a finite-slope graph can supply. At the default wall weight, several degrees into this regime, neural losses computed by quadrature are 1.6--1.8 times the bound from one cut per corner, with higher ratios occurring near the critical angle. Polynomial least-squares comparisons support the same geometric interpretation in a second smooth trial family. The bound supplies a physical scale for the residual compromise and explains why residual assessment must include wall flux and continuum compatibility.

\smallskip\par\noindent\textit{Short title: PINN approximations of capillary menisci}
\medskip

\section{Introduction}
Capillary menisci couple interface shape to wetting in channels, pores and low-gravity fluid systems \cite{r1,r3,r4,r5}. For example, pinned menisci in microfluidic devices serve as pressure-controlled interfaces \cite{rLabXu2007} and support acoustically driven capillary waves \cite{rLabXuPoF2007}. We consider the equilibrium shape of a fixed liquid volume in a cylindrical container without gravity. In square and equilateral-triangular cross-sections, the Concus--Finn critical contact angles are \(45^\circ\) and \(60^\circ\): below them an equilibrium graph cannot span the section, whereas above them an exact spherical patch is available \cite{r6,r7,r8}. Drop-tower experiments in a hexagonal cylinder \cite{r8} and orbital experiments and finite-element computations in double-proboscis containers \cite{r10} document related changes in liquid configuration near critical contact angles. For a numerical approximation, satisfying the curvature equation must therefore be assessed together with the contact-angle conditions that determine whether the interface can span the walls.

Physics-informed neural networks (PINNs) approximate solutions by minimizing residuals of governing equations and boundary conditions \cite{r12,r13}, with applications to the Young--Laplace equation \cite{r14}. Residual-based error estimates rely on stability of the continuum problem \cite{rMM}; finite collocation can leave distinct functions with the same sampled residuals \cite{r17}. Pseudo-time stepping with resampling can expose spurious approximations \cite{r16}. Urb\'an et al. also reported PINN approximations in parameter regions described as lacking mathematical solutions \cite{r32}. These observations motivate a quantitative question: when capillary geometry rules out equilibrium, what unavoidable cost does that incompatibility impose on the residual objective?

We express the classical capillary flux obstruction as a residual budget for smooth approximate interfaces. Finn's subdomain criterion provides the geometric deficit; a weighted estimate converts it into a certified lower bound on the continuous combined interior and wall loss for specified positive sampling densities and wall weight. The subdomain bounds and their sums over disjoint regions tend uniformly to zero as the contact angle approaches the threshold from below at fixed wall weight. This limits the diagnostic even when the continuum problem remains incompatible. Independently evaluated PINNs and polynomial least-squares fits relate the bound to attained residuals in two smooth trial families. The resulting comparison connects the physical flux demand to the scale and location of the numerical mismatch.

\section{Physical problem and loss bound}
\subsection{Constrained menisci and the classical reference}
Let \(\Omega\) be the cross-section and \(z=u(x,y)\) the interface position along the container axis, independent of its spatial orientation. The outward wall normal is \(\nu\), and \(\kappa=2H\) is twice the mean curvature. The contact angle \(\gamma\) is measured through the liquid, with \(0<\gamma\le\pi/2\). The zero-gravity capillary problem is \cite{r7,r8}
\begin{equation}\label{eq:problem}
\nabla\cdot Tu=\kappa,\qquad \nu\cdot Tu=\cos\gamma,
\qquad Tu=\frac{\nabla u}{\sqrt{1+|\nabla u|^2}}.
\end{equation}
Compatibility fixes \(\kappa=|\partial\Omega|\cos\gamma/|\Omega|\). The container length and liquid volume keep the interface clear of the end walls. We set \(u(0)=0\) to describe its shape; an axial translation fixes the volume. We seek a static, volume-constrained equilibrium meniscus in a finite container.

Scale lengths by the inradius \(L\) of the regular \(n\)-gon, so \(\kappa=2\cos\gamma\). With surface tension \(\sigma\), the dimensional curvature is \(\kappa/L\), and the Laplace pressure jump across the meniscus is \(\Delta P=P_{\rm gas}-P_{\rm liquid}=\sigma\kappa/L=2\sigma\cos\gamma/L\). The Bond number is \(\mathrm{Bo}=\Delta\rho\,gL^2/\sigma\), where \(\Delta\rho\) is the liquid--gas density difference and \(g\) is gravitational acceleration. We neglect gravity in the capillary-dominated regime \(\mathrm{Bo}\ll1\), as in microgravity experiments \cite{r10} or sufficiently small channels. Write \(\alpha=\pi/2-\pi/n\), half the corner angle. The classical Concus--Finn condition requires \(\gamma\ge\gamma_c=\pi/n\) for a spanning equilibrium graph \cite{r6,r7,r8}. The exact spherical meniscus solution on this side is the portion of a sphere
\begin{equation}\label{eq:cap}
u_{\rm cap}(x,y)=R-\sqrt{R^2-x^2-y^2},\qquad R=1/\cos\gamma.
\end{equation}
The sphere meets the vertical walls, giving the meniscus a polygonal footprint. At \(\gamma=\gamma_c\) its tangent is vertical at the vertices; the \(90^\circ\) limit is planar. Decreasing the contact angle strengthens wetting and increases the required wall flux until the spherical patch can no longer span the corners. We refer to \(\gamma<\gamma_c\) as the nonexistence regime and to \(\gamma>\gamma_c\) as the equilibrium regime, according to existence of the continuum graph.

\subsection{From capillary flux to residual loss}
\label{sec:fluxbound}\label{sec:corner}
Let \(p>0\) and \(q>0\) be interior and wall probability densities and \(w>0\) the wall weight. Define
\begin{equation}\label{eq:expectedloss}
\begin{split}
\mathcal L_{p,q}[u]&=\int_\Omega f^2p\,dA+w\oint_{\partial\Omega}g^2q\,ds,\\
f&=\nabla\cdot Tu-\kappa,\qquad g=\nu\cdot Tu-\cos\gamma.
\end{split}
\end{equation}
For a proper Lipschitz subdomain \(D\subset\Omega\), split its boundary into the internal cut \(\Sigma\) and wetted wall \(\Gamma\subset\partial\Omega\). Finn's classical subdomain criterion uses \cite{r7,r20,r21}
\begin{equation}\label{eq:finnfunctional}
\Phi(D)=|\Sigma|-\cos\gamma\,|\Gamma|+\kappa|D|.
\end{equation}
A negative value rules out a spanning equilibrium graph. The same flux balance constrains how closely a smooth trial graph can satisfy the curvature equation and wall condition together. The following weighted estimate expresses that constraint in the normalization of \eqref{eq:expectedloss}.

\emph{Proposition 1.} If \(B(D)=-\Phi(D)>0\), every smooth trial graph obeys
\begin{equation}\label{eq:subdomainloss}
\mathcal L_{p,q}[u]\ge
\frac{B(D)^2}{\displaystyle\int_Dp^{-1}\,dA+w^{-1}\int_\Gamma q^{-1}\,ds}.
\end{equation}
For disjoint subdomains with disjoint wetted walls, their right-hand sides can be summed. A finite denominator is required for a positive bound.

\emph{Proof.} The integrated residual balance and \(|Tu|\le1\) give \(\int_Df\,dA-\int_\Gamma g\,ds\ge B(D)\). Weighted Cauchy--Schwarz bounds the square of this expression by the local loss times the denominator in \eqref{eq:subdomainloss}; disjoint local losses sum to at most \(\mathcal L_{p,q}\). Here a smooth trial graph is any \(u\in C^2(\Omega)\cap C^1(\bar\Omega)\) with finite loss; networks with smooth activations and polynomials qualify.

Proposition~1 constrains the combined loss. Without control of the interior residual, a smooth graph can confine its wall mismatch to vertex neighborhoods of vanishing wall probability.

\begin{figure}[!htbp]
\centering
\includegraphics[width=4.8in]{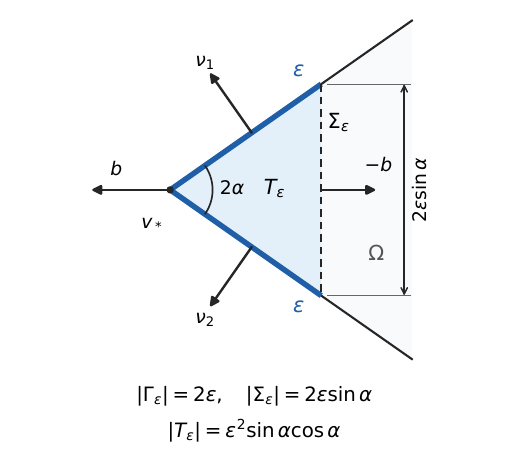}
\caption{Corner integration geometry in the channel cross-section. The vertex is \(v_*\); the two blue wetted-wall segments \(\Gamma_\varepsilon\) bound the shaded triangle \(T_\varepsilon\) and have total length \(2\varepsilon\). The artificial cut \(\Sigma_\varepsilon\) has length \(2\varepsilon\sin\alpha\), and \(|T_\varepsilon|=\varepsilon^2\sin\alpha\cos\alpha\). The unit normals \(\nu_1,\nu_2\) and their bisector \(b\) point outward from the polygon, whereas the outward normal of \(T_\varepsilon\) across \(\Sigma_\varepsilon\) is \(-b\), drawn on the cut. Thus \(\nu_1+\nu_2=2\sin\alpha\,b\).}
\label{fig:corner_geometry}
\end{figure}

For a polygon, choose the corner triangles \(T_\varepsilon\), with wetted walls \(\Gamma_\varepsilon\) and cut \(\Sigma_\varepsilon\), in Fig.~\ref{fig:corner_geometry}. Their flux deficit is
\[
B_\varepsilon=2\varepsilon\delta-\kappa\varepsilon^2\sin\alpha\cos\alpha,
\qquad \delta=\cos\gamma-\sin\alpha=\cos\gamma-\cos\gamma_c.
\]
The second form uses \(\sin\alpha=\cos(\pi/n)=\cos\gamma_c\), so \(\delta>0\) exactly in the nonexistence regime. For symmetric probability laws, the \(n\) disjoint corners give
\begin{equation}\label{eq:cornerloss}
\mathcal L_{p,q}[u]\ge\frac{nB_\varepsilon^2}{I_T(\varepsilon)+I_W(\varepsilon)/w},
\quad I_T=\int_{T_\varepsilon}p^{-1},\quad I_W=\int_{\Gamma_\varepsilon}q^{-1},
\end{equation}
whenever \(B_\varepsilon>0\). Equation~\eqref{eq:cornerloss} uses one cut at the same wall distance \(\varepsilon\) at each corner. Maximizing its right-hand side over admissible cutoffs gives the certified lower bound \(F\) in Table~\ref{tab:floor}. We call \(F\) the single-cut bound. Global flux balance places \(F\) strictly below the infimum over smooth graphs for the evaluation laws specified in Supplement Sec.~S2.

The corner also explains where a smooth approximation must depart from the wall condition. At the vertex, the two wall normals have bisector \(b\) and satisfy \(\nu_1+\nu_2=2\sin\alpha\,b\). Thus, in the nonexistence regime,
\begin{equation}\label{eq:vertexfloor}
\tfrac12(g_1+g_2)=\sin\alpha\,b\cdot Tu-\cos\gamma<-\delta,
\qquad \max(|g_1|,|g_2|)>\delta.
\end{equation}
The wall-flux demand exceeds what a finite slope can supply. Since \(|Tu|=|\nabla u|/(1+|\nabla u|^2)^{1/2}\), the slope is \(|\nabla u|=|Tu|/(1-|Tu|^2)^{1/2}\), so a vertex flux approaching its unit bound requires an unbounded slope. Equation~\eqref{eq:vertexfloor} holds pointwise at the vertex; a point carries no wall probability, so it does not by itself bound the wall RMS. Equation~\eqref{eq:cornerloss} measures the cost of distributing that deficit between interior and wall residuals over a neighborhood of the corner.

The bound becomes small near threshold. In an inradius-one regular polygon, for \(\gamma\le\gamma_c\), every subdomain bound satisfies
\begin{equation}\label{eq:uniformvanishing}
\frac{[B(D)]_+^2}{\int_Dp^{-1}+w^{-1}\int_\Gamma q^{-1}}
\le w\delta^2\int_\Gamma q\,ds\le w\delta^2.
\end{equation}
The same upper bound holds for sums over disjoint wetted walls. Hence even optimization over subdomains yields a bound tending uniformly to zero as \(\gamma\uparrow\gamma_c\), at fixed finite \(w\). Equation~\eqref{eq:uniformvanishing} is an upper envelope: the optimized corner bound decreases faster, as \(\delta^3\) under uniform laws and \(\delta^{8/3}\) under the clustered law (Supplement Sec.~S2). The estimate controls the scale of the certificate; attained losses remain determined by the trial field and numerical protocol. A short proof is given in Supplement Sec.~S2.

\section{Numerical protocol}
\label{sec:methods}
The \emph{Angle survey} uses four-hidden-layer tanh networks of width 48 in double precision, covering nine contact angles for the square cross-section and six for the triangular cross-section. We minimize the sampled version of \eqref{eq:expectedloss} plus \(u(0)^2\), with 3000 interior points, 800 wall points and default \(w=10\). Interior sampling mixes uniform area with corner clustering; wall sampling clusters points toward side endpoints. Training uses 2000 Adam steps followed by L-BFGS with at most 500 iterations and the stopping settings in Supplement Sec.~S1.

The \emph{Matched capacity} comparison uses the same optimization schedule at widths 48, 64 and 128, at contact angles \(44^\circ/46^\circ\) for the square cross-section and \(58^\circ/62^\circ\) for the triangular cross-section. \emph{Double collocation} doubles both point counts; \emph{Low wall weight} and \emph{High wall weight} use \(w=1\) and 100. A \emph{Width survey} in the square cross-section at contact angles \(40^\circ/50^\circ\) uses widths 16, 32, 64 and 128. Same-seed angle pairs share initialization and collocation. The 48 near-threshold pairs compare the nonexistence and equilibrium regimes and include reused seeds and point sets across widths.

We evaluate the mean-square residuals by deterministic Gauss quadrature under the clustered evaluation laws (Supplement Secs.~S2 and S6). We report normalized interior and wall RMS,
\[
R_i=\sqrt{\mathbb E_p f^2}/\kappa,\qquad
R_w=\sqrt{\mathbb E_q g^2}/\cos\gamma,
\]
along with the expected combined loss. Evaluation under uniform area and arclength measures checks sensitivity to the choice of measure. Supplement Sec.~S1 describes the local wall probes and vertex flux diagnostics, and Sec.~S4 gives the polynomial least-squares comparison.

\paragraph*{Research ethics.}
This computational study involved no human participants or animals; research ethics approval and informed consent were not required.

\paragraph*{Computational assistance.}\label{sec:verify}
Claude (Anthropic) and Codex (OpenAI) assisted with code development, numerical execution, data analysis and plotting under the authors' direction. Outputs were checked against the exact spherical meniscus solution, independently evaluated residuals, deterministic quadrature and capillary flux balances.

\section{Results}
\subsection{Small bulk residuals and a corner wall-flux deficit}
\label{sec:spurious}\label{sec:exists}
\begin{figure}[!htbp]
\centering
\includegraphics[width=\linewidth]{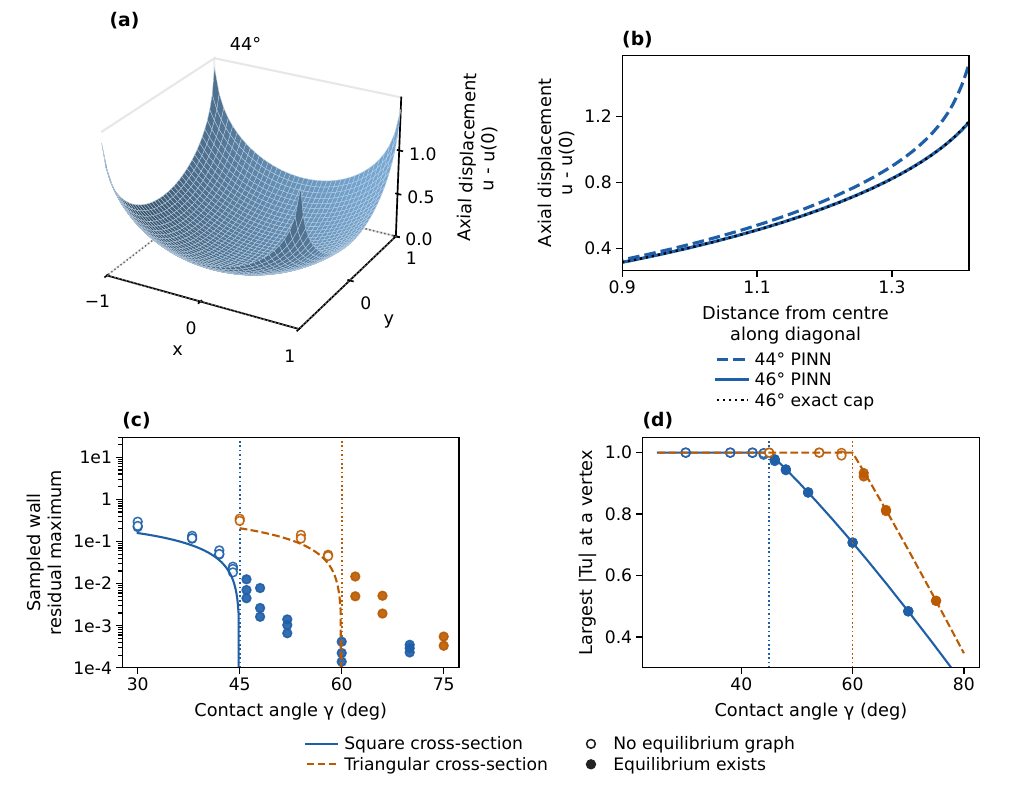}
\caption{Computed surface shape, near-corner profiles and corner diagnostics from the width-48 Angle survey. (a) The seed-0 PINN graph in the square cross-section at \(44^\circ\), plotted over the full cross-section despite nonexistence of a spanning equilibrium. (b) Its diagonal profile and the same-seed \(46^\circ\) profile, compared with the exact \(46^\circ\) cap over \(0.9\le r\le\sqrt{2}\). Lengths are scaled by the inradius and \(u-u(0)\) is axial displacement. (c) Largest sampled unnormalized wall residual \(|g|=|\nu\cdot Tu-\cos\gamma|\) in the local corner probes, not the normalized RMS \(R_w\), and (d) largest vertex flux magnitude \(|Tu|\). In (c,d), circles show individual PINN runs: seeds 0--2 for the square cross-section and 0--1 for the triangular cross-section. Theoretical reference curves show \(\delta=\cos\gamma-\cos\gamma_c\) in the nonexistence regime in (c), and \(\min\{1,\cos\gamma/\sin\alpha\}\) in (d). Dotted vertical lines mark \(\gamma_c=45^\circ\) and \(60^\circ\).}
\label{fig:profiles}
\end{figure}

In the equilibrium regime (\(\gamma>\gamma_c\)), the networks approximate the exact spherical patch. In the nonexistence regime (\(\gamma<\gamma_c\)), every evaluated strong-form run returns a smooth field with finite axial extent at its prescribed budget. Near the critical angle, the interior residuals in both regimes are of order \(10^{-3}\)--\(10^{-2}\) of the prescribed curvature. Figure~\ref{fig:profiles} shows a complete computed graph, its near-corner profile and the distinct corner behavior despite similar bulk errors.

In the nonexistence regime, with \(w\ge10\), the vertex flux approaches its unit bound while the adjoining contact-angle conditions demand more. The sampled wall-residual maximum exceeds the vertex bound \eqref{eq:vertexfloor} in every such network; the sampling of these maxima is specified in Supplement Sec.~S1. In the equilibrium regime, the exact vertex flux is \(\cos\gamma/\sin\alpha<1\). The physical distinction is therefore located at the wall and its corners even when the bulk equations are approximated well in both regimes.

\begin{figure}[!htbp]
\centering
\includegraphics[width=\linewidth]{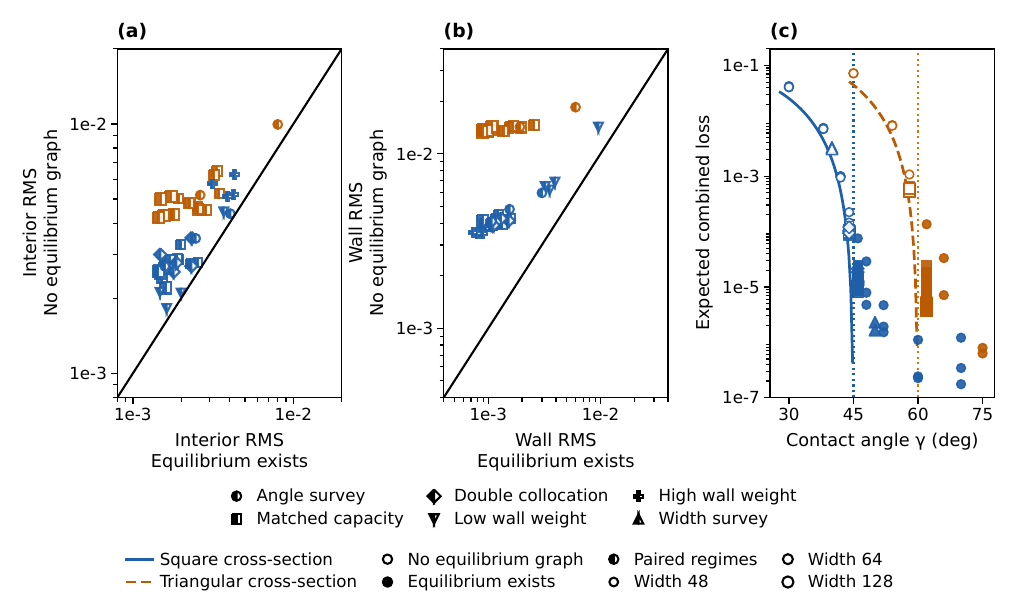}
\caption{Paired residuals and expected combined loss. (a) Interior RMS in the nonexistence regime (\(44^\circ\) for the square cross-section, \(58^\circ\) for the triangular cross-section) against the same seed and width in the equilibrium regime (\(46^\circ\), \(62^\circ\)), for all 48 matched pairs; (b) the corresponding wall RMS. Black lines mark equality. (c) Expected combined loss at \(w=10\). All plotted RMS and losses use Gauss order 128 under the clustered evaluation laws (Supplement Sec.~S6). In (c), the curves show the certified lower bounds and the dotted vertical lines mark \(\gamma_c\). Protocols and weights are given in Supplement Table~S1.}
\label{fig:paired_residuals}
\end{figure}

Figure~\ref{fig:paired_residuals} compares the same-seed pairs. All 48 have larger interior and wall RMS in the nonexistence regime. Interior RMS ratios range from 1.05 to 3.36, and wall and training-loss ratios are generally larger. Same-regime comparisons also show substantial changes (Supplement Table~S3).

\begin{samepage}
Across widths, the mean wall residual in the equilibrium regime decreases by about 30 percent from 48 to 128, while the values in the nonexistence regime remain near a plateau. This plateau is an observation at the tested widths and optimization budget. Proposition~1 bounds the combined loss and does not imply a positive lower bound on the wall RMS alone (Sec.~\ref{sec:fluxbound}); the plateau reflects how the trained fields divide the flux deficit between the two terms. At \(w=10\) the wall term carries 79--99 percent of the quadrature loss of the fields in Table~\ref{tab:floor}. Uniform-measure checks retain the close bulk-residual scales and show how they depend on the evaluation measure (Supplement Table~S2).
\par
\end{samepage}

\newpage
\subsection{Attained loss relative to the geometric bound}
Table~\ref{tab:floor} compares quadrature-evaluated expected losses with the certified single-cut bound \(F\) of \eqref{eq:cornerloss}. At the default wall weight, the PINN fits several degrees below the thresholds attain losses 1.57--1.83 times \(F\). Thus more than half of each attained loss is unavoidable for smooth graphs. The infimum remains undetermined. At the closest tested angles below the thresholds, the ratio reaches about 7.2 in the square and 3.4 in the triangle.

In the square cross-section, the polynomial fits support the same interpretation of the residual scale (Supplement Sec.~S4). For the cases farther below threshold, two nested corner cuts yield a stronger certified lower bound (Supplement Sec.~S2).

\begin{table}[!htbp]
\begin{minipage}{\linewidth}
\setlength{\belowcaptionskip}{4pt}
\caption{Certified lower bound \(F\) and attained expected loss \(\mathcal L\). Both use the clustered evaluation law and the stated wall weight. \(F\) is the single-cut bound \eqref{eq:cornerloss}, and \(\mathcal L\) is the combined loss evaluated by quadrature (Supplement Sec.~S6). \(F_{\rm unif}/F\) is the uniform-to-clustered bound ratio, and \(\mathcal L/F\) is the loss-to-bound ratio. \(\delta=\cos\gamma-\cos\gamma_c\). Ranges span the runs. The \(40^\circ\) square fits use width 128.}
\label{tab:floor}
\small
\begingroup\setlength{\tabcolsep}{3pt}\renewcommand{\arraystretch}{1.15}
\begin{tabular*}{\linewidth}{@{\extracolsep{\fill}}lcccccccc@{}}
\toprule
\shortstack{Cross-\\section} & \(\gamma\) & \(w\) & \(\delta\) & \(F\) & \(F_{\rm unif}/F\) & Runs & \(\mathcal L\) & \(\mathcal L/F\) \\
\midrule
Square & $30^\circ$ & 10 & 0.1589 & $2.53\times10^{-2}$ & 0.473 & 3 & $4.05\text{--}4.24\times10^{-2}$ & 1.60--1.67 \\
Square & $38^\circ$ & 10 & 0.0809 & $4.50\times10^{-3}$ & 0.408 & 3 & $7.12\text{--}7.45\times10^{-3}$ & 1.58--1.66 \\
Square & $40^\circ$ & 10 & 0.0589 & $1.97\times10^{-3}$ & 0.377 & 2 & $3.11\text{--}3.32\times10^{-3}$ & 1.57--1.68 \\
Square & $42^\circ$ & 10 & 0.0360 & $5.44\times10^{-4}$ & 0.330 & 3 & $9.44\text{--}9.97\times10^{-4}$ & 1.74--1.83 \\
Square & $44^\circ$ & 10 & 0.0122 & $3.12\times10^{-5}$ & 0.239 & 23 & $0.89\text{--}2.24\times10^{-4}$ & 2.85--7.18 \\
Square & $44^\circ$ & 1 & 0.0122 & $3.12\times10^{-6}$ & 0.242 & 4 & $0.27\text{--}1.46\times10^{-4}$ & 8.59--46.76 \\
Square & $44^\circ$ & 100 & 0.0122 & $3.11\times10^{-4}$ & 0.213 & 4 & $6.92\text{--}7.16\times10^{-4}$ & 2.22--2.30 \\
Triangular & $45^\circ$ & 10 & 0.2071 & $4.44\times10^{-2}$ & 0.455 & 2 & $7.217\text{--}7.224\times10^{-2}$ & 1.62--1.63 \\
Triangular & $54^\circ$ & 10 & 0.0878 & $5.18\times10^{-3}$ & 0.392 & 2 & $8.13\text{--}8.36\times10^{-3}$ & 1.57--1.61 \\
Triangular & $58^\circ$ & 10 & 0.0299 & $3.16\times10^{-4}$ & 0.302 & 17 & $0.52\text{--}1.07\times10^{-3}$ & 1.63--3.39 \\
\bottomrule
\end{tabular*}\endgroup
\end{minipage}
\end{table}

The single-cut bound decreases rapidly near the critical angle, consistent with \eqref{eq:uniformvanishing}. In the square, it becomes comparable to losses of near-critical fits in the equilibrium regime. Small expected losses in the nonexistence regime coexist with localized wall deficits. Changing the wall weight changes both the attained loss and the certified lower bound (Table~\ref{tab:floor}).

The rounded-square fits show large residual peaks on short arcs, including in the cases where the existence of an equilibrium graph was left unclassified. The integrated deficit certifies nonexistence at the smallest angle tested (Supplement Sec.~S5).

\newpage
\section{Discussion and conclusions}\label{sec:discussion}

For the static meniscus problem posed as a single graph, corner flux balance gives a lower bound on the combined loss even when the interior curvature residual is small. The bound weakens as the contact angle approaches its critical value from the nonexistence regime. If liquid reaches the end walls of a finite container, the equilibrium model must also account for the conditions at those walls \cite{r10}.

A neural Ritz formulation minimizes energy over trial graphs represented by a neural network \cite{rDeepRitz,rDEM}. The classical pressure-augmented capillary energy \cite{r21} takes the form
\begin{equation}\label{eq:energy}
E_\kappa[u]=\int_\Omega\sqrt{1+|\nabla u|^2}\,dA-\cos\gamma\oint_{\partial\Omega}u\,ds+\kappa\int_\Omega u\,dA.
\end{equation}
Below the critical angle, a corner-rise direction makes this continuum energy unbounded below over graphs of unrestricted height, even at fixed volume (Supplement Sec.~S2). A positive flux deficit rules out a spanning equilibrium regardless of the numerical formulation. In the existence regime, the Ritz objective uses first derivatives and incorporates wetting through the wall term. Strong-form collocation penalizes curvature and wall residuals separately, which helps locate mismatches.

Variational PINNs (VPINNs) use test functions to measure weak-form residuals \cite{rVPINN}. A weak formulation that retains the wall term detects the same deficit when its test space contains a function \(\varphi\) with \(J(\varphi)<0\), where \(J\) is defined in Supplement Sec.~S2. In the original VPINN examples, the test functions vanish on the boundary and a separate penalty accounts for the boundary condition.

The deficit also bears on the corner slope. As the magnitude of the mean of the two one-sided vertex residuals in Eq.~\eqref{eq:vertexfloor} approaches \(\delta\), \(b\cdot Tu\to1\), which requires an unbounded slope. All measured slopes are finite. The largest slopes grow with polynomial degree at each tested angle below \(45^\circ\), while the neural trends depend on the cross-section (Supplement Tables~S8 and S9). These tests leave unresolved whether the corner slope diverges and whether the centered heights \(u-u(0)\) converge in \(L^2(\Omega)\) as capacity increases.

A practical assessment should combine independent interior and wall residuals, local flux probes and an integrated balance. Probes locate a mismatch; the certificate tests its compatibility. In the rounded-section example, the positive deficit certifies nonexistence, while local arc peaks also occur at unclassified comparison angles. Resampling changes the collocation points. Pseudo-time stepping uses a relaxed interior residual that depends on the previous iterate \cite{r16}. Corner weighting changes where residuals are penalized most. The comparison uses the original loss in Eq.~\eqref{eq:expectedloss} and the bound evaluated under the same sampling densities and wall weight.

Adaptive loss weighting, such as learning-rate annealing \cite{rLRA}, GradNorm \cite{rGradNorm} or ReLoBRaLo \cite{rReLoBRaLo}, changes the cost assigned to the two residuals. At each choice of positive weights, Proposition~1 applies to any subdomain with \(B(D)>0\). It does not predict weight divergence or stiffness. Evaluating the certificate with the final weights helps interpret a persistent wall residual.

\paragraph*{Conflict of interest}
The authors state no conflict of interest.

\newpage
\paragraph*{Author contributions}
All authors have accepted responsibility for the entire content of this manuscript and approved its submission.

\textbf{Zijian Liu:} Formal analysis (equal); Investigation (equal); Methodology (equal); Software (equal); Writing -- original draft (equal); Writing -- review \& editing (equal).

\textbf{Julian Olszewski:} Formal analysis (equal); Funding acquisition (lead); Investigation (equal); Methodology (equal); Software (equal); Writing -- original draft (equal); Writing -- review \& editing (equal).

\textbf{Jie Xu:} Conceptualization; Supervision (lead); Writing -- review \& editing (equal).

\paragraph*{Research funding}
J.~Olszewski acknowledges support from the University of Illinois Chicago Chancellor's Undergraduate Research Award (CURA).

\paragraph*{Data availability}
The code and data supporting this study are openly available in the University of Illinois Chicago repository INDIGO \mbox{(\url{https://doi.org/10.25417/uic.34989052})}.

\paragraph*{Acknowledgments}
Claude (Anthropic) and Codex (OpenAI) assisted with code development, supported scientific discussion and provided editorial assistance with the manuscript under the authors' direction. The authors reviewed and edited this material and are responsible for the final interpretation and manuscript.

\FloatBarrier

\clearpage
\setcounter{section}{0}
\setcounter{subsection}{0}
\setcounter{equation}{0}
\setcounter{table}{0}
\setcounter{figure}{0}
\renewcommand{\thesection}{S\arabic{section}}
\renewcommand{\thesubsection}{\thesection.\arabic{subsection}}
\makeatletter\renewcommand{\p@subsection}{}\makeatother
\renewcommand{\thetable}{S\arabic{table}}
\renewcommand{\thefigure}{S\arabic{figure}}
\renewcommand{\theequation}{S\arabic{equation}}

\begin{center}
{\Large\bfseries Supplementary material: Physics-informed neural network (PINN) approximations of capillary menisci in polygonal channels\par}
\vspace{0.8em}
Zijian Liu, Julian Olszewski, and Jie Xu\textsuperscript{*}\\[0.5em]
{\small Department of Mechanical and Industrial Engineering, University of Illinois Chicago,\\ 842 W. Taylor Street, Chicago, IL 60607, USA}
\end{center}
\begingroup
\renewcommand{\thefootnote}{}
\footnotetext[0]{Zijian Liu and Julian Olszewski contributed equally to this work.}
\endgroup
\begingroup
\renewcommand{\thefootnote}{\fnsymbol{footnote}}
\footnotetext[1]{Corresponding author: jiexu@uic.edu.}
\endgroup
\vspace{0.5em}

This supplement gives the numerical protocols, sampling-law integrals, measure comparisons, polynomial fits, rounded-domain geometry and the deterministic quadrature of expected losses supporting the main text. We use the main-text notation: \(u-u(0)\) is an axial displacement, \(\kappa=2H\), and \(f,g\) are the interior and wall residuals of Eq.~3. For the sharp polygons, the nonexistence regime is \(\gamma<\gamma_c\), where no spanning equilibrium graph exists; the equilibrium regime is \(\gamma>\gamma_c\), where the exact cap is available.

\section{Neural protocols and diagnostics}
\label{supp:training}\label{supp:methods}
All neural fields are fully connected networks with two spatial inputs, four hidden tanh layers and one linear output, using double precision. Unless varied in Table~\ref{tab:populations}, the hidden-layer width is 48, the fixed collocation set has \(N=3000\) interior and \(M=800\) wall points, and \(w=10\). Weights between layers of sizes \(a,b\) are independent normal variates with standard deviation \(\sqrt{2/(a+b)}\); biases are zero. The objective is
\begin{equation}\label{eq:empiricalloss}
 \widehat{\mathcal L}=\frac1N\sum_{i=1}^{N}f(x_i)^2+
 \frac{w}{M}\sum_{j=1}^{M}g(y_j)^2+u(0)^2.
\end{equation}
The last term fixes axial translation and is separate from the continuous residual objective in the main text.

Training uses 2000 Adam steps at initial learning rate \(10^{-3}\), multiplied by 0.8 every 750 steps, followed by L-BFGS with a strong Wolfe line search. L-BFGS uses learning rate 1, history size 50, at most 500 iterations and 625 closure evaluations, gradient tolerance \(10^{-12}\), and change tolerance \(10^{-14}\). The Angle survey and width-16/32/64 Width survey use one CPU thread; the width-128 Width survey and the other groups use an NVIDIA GeForce RTX 3070 Laptop GPU (PyTorch 2.11.0, CUDA 12.8). All reported runs complete the optimizer calls under the same stated schedule and stopping criteria.

The Angle survey covers contact angles \(30,38,42,44,46,48,52,60,70\) degrees for the square cross-section (seeds 0--2) and \(45,54,58,62,66,75\) degrees for the triangular cross-section (seeds 0--1). The remaining groups are defined in Table~\ref{tab:populations}. For a given seed, width and implementation, initialization and collocation are generated before the contact angle enters. Same-seed angle pairs therefore share both; widths reuse the same seed and points. The 48 near-threshold angle pairs in the main text are descriptive paired comparisons, not independent replications.

\begin{table}[!htbp]
\begin{minipage}{\linewidth}
\small
\caption{Neural groups. Square (S) and triangular (T) cross-sections have inradius one. The counts \(N,M\) refer to interior and wall collocation points; all groups use \(w=10\) except the two weight variations. Angle lists for the Angle survey are given in the text; the S and T pairs are respectively \(44^\circ/46^\circ\) and \(58^\circ/62^\circ\).}
\label{tab:populations}

\begingroup\setlength{\tabcolsep}{3pt}\renewcommand{\arraystretch}{1.15}
\begin{tabular*}{\linewidth}{@{\extracolsep{\fill}}lclcccc@{}}
\toprule
Group & width & \shortstack{Cross-section\\and angles} & seeds & \(N/M\) & device & runs \\
Angle survey & 48 & S and T, listed angles & 0--2 / 0--1 & 3000/800 & CPU & 39 \\
Width survey & 16,32,64 & S, \(40^\circ/50^\circ\) & 0--1 & 3000/800 & CPU & 12 \\
Matched capacity & 48,64,128 & S and T pairs & 3--7 & 3000/800 & GPU & 60 \\
Double collocation & 48 & S pair & 3--7 & 6000/1600 & GPU & 10 \\
Low wall weight, \(w=1\) & 48 & S pair & 1--4 & 3000/800 & GPU & 8 \\
High wall weight, \(w=100\) & 48 & S pair & 1--4 & 3000/800 & GPU & 8 \\
Width survey & 128 & S, \(40^\circ/50^\circ\) & 0--1 & 3000/800 & GPU & 4 \\
\bottomrule
\end{tabular*}\endgroup
\end{minipage}
\end{table}

The evaluation law uses an interior clustering scale of 0.2 (Sec.~\ref{app:floor}). Deterministic Gauss quadrature gives the mean-square residuals (Sec.~\ref{supp:quadrature}). We report
\begin{equation}
 R_i=\sqrt{\mathbb E_p f^2}/\kappa,
 \qquad R_w=\sqrt{\mathbb E_q g^2}/\cos\gamma.
\end{equation}
Training losses use the fixed collocation points (Eq.~\ref{eq:empiricalloss}). Vertex fluxes use exact network derivatives. The CPU Angle and Width surveys probe the local wall maximum at 40 points on each adjoining wall, at distances from \(10^{-6}\) to \(10^{-3}\) from each vertex. GPU groups use the maximum over the independent wall probes of Sec.~\ref{supp:quadrature}. These maxima characterize the residual at the probe points.

Cap error is the RMS of \([u-u(0)]-u_{\rm cap}\), divided by the centered RMS of \(u_{\rm cap}\), under the stated evaluation measure. Uniform-area checks divide the polygon into triangles from its center and apply a Duffy map with tensor Gauss--Legendre rules of orders 64 and 128. Uniform-arclength checks split each wall into two endpoint intervals of length 0.03 and the intervening central interval, using orders 256 and 512 on each. The finer rules give the reported geometric-measure values. GPU checkpoints are independently reevaluated on the CPU using the same residual and flux definitions.

\section{Sampling densities and loss-bound evaluation}
\label{app:floor}\label{supp:bound}
We use the capillary flux and Finn functional defined in the main text \cite{supp-r6,supp-r7,supp-r20}. This section gives the sampling densities needed to evaluate the weighted bound.

For an inradius-one regular \(n\)-gon, write \(\alpha=\pi/2-\pi/n\), side length \(\ell=2\tan(\pi/n)\), area \(A=n\tan(\pi/n)\), and perimeter \(P=n\ell\). The interior sampler mixes probability \(3/4\) uniform area with probability \(1/4\) near vertices. A clustered point chooses a vertex and an inward wedge direction uniformly, then takes radius \(r=r_0+sU^2\), where \(r_0=10^{-4}\) and \(U\) is uniform on \([0,1]\). Here \(s=0.25\) in training and \(s=0.2\) in evaluation. Near a vertex its area density is
\begin{equation}\label{eq:pdensity}
 p_s(r)=\frac{3}{4A}+c_s(r),\qquad
 c_s(r)=\begin{cases}
 [16n\alpha r\sqrt{s(r-r_0)}]^{-1},&r_0<r<r_0+s,\\
 0,&\text{otherwise}.
 \end{cases}
\end{equation}
Away from the disjoint vertex-cluster supports only the uniform term remains. Wall sampling chooses a side uniformly and maps \(z\), uniform on \([-1,1]\), to the side coordinate
\[
 t=\operatorname{sign}(z)\bigl[1-(1-|z|)^{3/2}\bigr].
\]
At arclength \(0<d\le\ell/2\) from the nearest endpoint, its probability density is
\begin{equation}\label{eq:qdensity}
 q(d)=\frac{2}{3n\ell}\left(\frac{2d}{\ell}\right)^{-1/3}.
\end{equation}

For a corner triangle cut at wall distance \(\varepsilon\le\ell/2\), define \(I_T=\int_{T_\varepsilon}p^{-1}\,dA\) and \(I_W=\int_{\Gamma_\varepsilon}q^{-1}\,ds\). The inverse-density integrals are
\begin{align}
 I_T&=\int_{-\alpha}^{\alpha}\int_0^{\varepsilon\cos\alpha/\cos\theta}
       \frac{r}{p_s(r)}\,dr\,d\theta,\\
 I_W&=\frac{9n\ell^2}{8}\left(\frac{2\varepsilon}{\ell}\right)^{4/3}.
\end{align}
For uniform probability measures they reduce to \(I_T=A|T_\varepsilon|\) and \(I_W=2\varepsilon P\). Adaptive quadrature evaluates \(I_T\), splitting at \(r_0\); a bracketed scalar search maximizes the main-text corner bound over positive-deficit disjoint triangles. For the cases and weights of Table~1 the training- and evaluation-law bounds agree within 0.3 percent; over all cached combinations the largest difference is 1.2 percent, at \(w=100\). 

For the symmetric corner family,
\[
 B_\varepsilon=2\varepsilon\delta-H\varepsilon^2\sin2\alpha,
 \qquad \delta=\cos\gamma-\sin\alpha=\cos\gamma-\cos\gamma_c.
\]
At fixed finite \(w\) and \(\delta\to0^+\), the maximizing cutoff is proportional to \(\delta\). The denominator is then dominated by \(I_W/w\), yielding \(\delta^3\) for uniform laws and \(\delta^{8/3}\) for the clustered law. Relative to \(\varepsilon^*=\delta/(H\sin2\alpha)\), the maximizing cutoff tends to \(2/3\) for uniform laws and \(1/2\) for the clustered law. At \(w=10\) the computed clustered-law cutoffs are 0.47--0.50 times \(\varepsilon^*\) for the angles of Table~1.

To check uniform vanishing over all subdomains, take \(\gamma\le\gamma_c\) and the feasible critical field \(\sigma_c=\cos\gamma_c(x,y)\). Since \(P=2A\),
\[
 B_\gamma(D)=-\Phi_{\gamma_c}(D)+\delta(|\Gamma|-2|D|)
 \le\delta|\Gamma|.
\]
The inequality \(|\Gamma|^2\le(\int_\Gamma q^{-1})(\int_\Gamma q)\) gives Eq.~8. Disjoint wetted-wall masses sum to at most one, so optimizing or summing these certificates still gives an upper bound \(w\delta^2\), which vanishes as \(\gamma\to\gamma_c^-\) at fixed \(w\).

\emph{Complement bound.} For a region \(U\) with wetted wall \(\Gamma_U\), define
\[
 \mathcal R(U)=\int_U f\,dA-\int_{\Gamma_U}g\,ds,\qquad
 d_U=\int_U p^{-1}\,dA+w^{-1}\int_{\Gamma_U}q^{-1}\,ds.
\]
Let \(B_U=B(U)\) as in Proposition~1, and let \(\mathcal L_U\) be the loss on \(U\) and \(\Gamma_U\).
The divergence theorem and \(\kappa|\Omega|=\cos\gamma\,|\partial\Omega|\) give \(\mathcal R(\Omega)=0\). Thus the complement \(U^c=\Omega\setminus U\), with the remaining wetted wall, has \(\mathcal R(U^c)=-\mathcal R(U)\).
For \(\mathcal R(U)\ge B_U>0\) and finite positive \(d_U,d_{U^c}\), weighted Cauchy--Schwarz on the two regions gives
\[
 \mathcal L_{p,q}[u]=\mathcal L_U+\mathcal L_{U^c}
 \ge B_U^2\left(\frac{1}{d_U}+\frac{1}{d_{U^c}}\right).
\]
For the union of the \(n\) disjoint corner triangles, \(B_U=nB_\varepsilon\). At the cutoff maximizing the single-cut bound \(F\), the first term equals \(F\) and the second is positive. Therefore \(F\) lies strictly below the infimum for the laws used here.

\emph{Two nested corner cuts.} Let \(U_1\subset U_2\) be the unions of the \(n\) corner triangles at cutoffs \(\varepsilon_1<\varepsilon_2\), with positive deficits \(\beta_j=nB_{\varepsilon_j}\) and \(d_j=d_{U_j}\).
For \(c_1,c_2\ge0\), adding the flux inequalities gives \(c_1\mathcal R(U_1)+c_2\mathcal R(U_2)\ge c_1\beta_1+c_2\beta_2\). Applying weighted Cauchy--Schwarz to the resulting weight, which is \(c_1+c_2\) on \(U_1\) and \(c_2\) on \(U_2\setminus U_1\), and to the corresponding wall weight gives
\[
 \mathcal L_{p,q}[u]\ge
 \frac{(c_1\beta_1+c_2\beta_2)^2}{d_1(c_1+c_2)^2+(d_2-d_1)c_2^2}.
\]
When \(c_2=(\beta_2-\beta_1)/(d_2-d_1)\) and \(c_1=\beta_1/d_1-c_2\) are nonnegative, they maximize this bound for the two fixed cuts and yield
\[
 F_2=\frac{\beta_1^2}{d_1}+\frac{(\beta_2-\beta_1)^2}{d_2-d_1}.
\]
For Table~1 we set \(\varepsilon_0=2\varepsilon^*=2\delta/(\kappa\sin\alpha\cos\alpha)\), the positive zero of \(B_\varepsilon\), and choose
\(\varepsilon_1=(9/50)^{3/2}\varepsilon_0\) and \(\varepsilon_2=(23/50)^{3/2}\varepsilon_0\).
Both coefficients are positive in every row, and the triangles at different vertices are disjoint at each cutoff. Using Eqs.~(S5)--(S6) under the evaluation law gives \(F_2/F=1.207\text{--}1.209\), while the complement bound reaches about \(1.011F\). The loss ratios \(\mathcal L_{p,q}[u]/F_2\) are 1.30--1.52 for the 15 farther-subcritical neural fits at \(w=10\). The infimum remains undetermined.

\emph{Wall term alone.} Outside intervals of length \(\eta\) at the endpoints of each side, let \(u\) vanish on the wall with outward normal derivative \(\cot\gamma\) and zero tangential derivative. Then \(\nu\cdot Tu=\cos\gamma\) there. Blending these edge collars smoothly to zero near the vertices confines the wall mismatch to intervals of total wall probability \((2\eta/\ell)^{2/3}\) by Eq.~\ref{eq:qdensity}. Since \(|g|\le1+\cos\gamma\), the wall loss tends to zero.
For a fixed subdomain \(D\) with \(B(D)>0\), let \(a=\int_D pf^2\,dA\), \(I_D=\int_Dp^{-1}\,dA\), and \(I_\Gamma=\int_\Gamma q^{-1}\,ds\), with finite positive \(I_D,I_\Gamma\). The integrated balance gives
\begin{equation}\label{eq:conditionalwall}
 \int_\Gamma qg^2\,ds\ge\frac{\bigl[B(D)-(aI_D)^{1/2}\bigr]_+^2}{I_\Gamma}.
\end{equation}
The right-hand side is positive for \(a<B(D)^2/I_D\). If the wall loss tends to zero, the lower limit of the interior loss on \(D\) is at least \(B(D)^2/I_D\).

\emph{Energy and tested residuals.} The first variation of the pressure-augmented energy in main-text Eq.~(9) gives the curvature and wall equations. Without the pressure term or the volume constraint, translation by \(C\) changes the energy by \(-C|\partial\Omega|\cos\gamma\). For the augmented energy and a smooth direction \(\varphi\),
\[
 \lim_{t\to\infty} E_\kappa[u+t\varphi]/t
 =J(\varphi)=\int_\Omega|\nabla\varphi|\,dA
 -\cos\gamma\oint_{\partial\Omega}\varphi\,ds+\kappa\int_\Omega\varphi\,dA.
\]
Smoothing an indicator across the artificial cut gives \(J(\varphi)\to\Phi(D)\): the variation term tends to the cut length and the wall and area terms tend to those for \(D\). A corner cut with \(\Phi(D)<0\) therefore gives an energy-decreasing direction along which the energy tends to \(-\infty\). Subtracting the mean of \(\varphi\) preserves the volume and leaves \(J\) unchanged by flux compatibility.

For the natural-boundary weak formulation, integration by parts gives
\[
 r(\varphi;u)=\oint_{\partial\Omega}g\varphi\,ds-\int_\Omega f\varphi\,dA
 =\int_\Omega Tu\cdot\nabla\varphi\,dA+\kappa\int_\Omega\varphi\,dA
   -\cos\gamma\oint_{\partial\Omega}\varphi\,ds.
\]
Since \(|Tu|\le1\), \(r(\varphi;u)\le J(\varphi)\). A test with \(J(\varphi)<0\) has a nonzero tested residual for every trial. For tests with zero wall trace, \(r\) contains only the interior residual and the contact angle must be assessed separately.

\section{Evaluation measures and same-side comparisons}
\label{app:measures}\label{supp:comparisons}
Table~\ref{tab:measures} gives the Angle-survey results. Under uniform measures, the interior-RMS ranges overlap across the paired angles, and the wall-RMS ranges remain separated.

\begin{table}[!htbp]
\begin{minipage}{\linewidth}
\small
\caption{Seed ranges of the Angle survey under uniform and clustered quadrature. All entries are in units of \(10^{-3}\). Interior RMS is divided by \(2H\), wall RMS by \(\cos\gamma\); cap error is the relative error under the stated area or corner-weighted measure. No cap reference exists for the nonexistence-regime rows.}
\label{tab:measures}\label{tab:geometric}
\begingroup\setlength{\tabcolsep}{3pt}\renewcommand{\arraystretch}{1.15}
\begin{tabular*}{\linewidth}{@{\extracolsep{\fill}}lccccc@{}}
\toprule
\shortstack{Cross-\\section} & \(\gamma\) & \shortstack{Interior RMS\\(area)} & \shortstack{Wall RMS\\(arclength)} & \shortstack{Cap error\\(area)} & \shortstack{Cap error\\(corner weighted)} \\
square & $44^\circ$ & $[1.96, 4.20]$ & $[2.09, 3.34]$ & n/a & n/a \\
square & $46^\circ$ & $[1.39, 4.21]$ & $[0.647, 1.87]$ & $[0.498, 2.05]$ & $[3.80, 13.7]$ \\
triangular & $58^\circ$ & $[3.97, 10.6]$ & $[7.36, 10.9]$ & n/a & n/a \\
triangular & $62^\circ$ & $[2.67, 7.75]$ & $[1.34, 4.00]$ & $[1.24, 5.08]$ & $[2.66, 13.8]$ \\
\bottomrule
\end{tabular*}\endgroup
\end{minipage}
\end{table}

Table~\ref{tab:sameside} gives the same-seed Angle-survey ratios for angle pairs within one regime and across the two regimes. For the square cross-section, paired contact angles differ by two degrees, except the four-degree 48/52 pair; both pairs for the triangular cross-section differ by four degrees. Large ratios also occur within a single regime; these descriptive comparisons do not establish a discontinuity or a universal residual cutoff.

\begin{table}[!htbp]
\begin{minipage}{\linewidth}
\small
\caption{Angle-survey ratios at lower angle divided by those at higher angle. Ranges cover seeds 0--2 for the square cross-section and 0--1 for the triangular cross-section. Interior and wall RMS use Gauss order 128 under the clustered evaluation law. \(\widehat{\mathcal L}\) is the fixed-collocation training loss. N denotes the nonexistence regime (\(\gamma<\gamma_c\), no spanning equilibrium graph) and E the equilibrium regime (\(\gamma>\gamma_c\), exact cap available), with \(\gamma_c=45^\circ\) for the square cross-section and \(60^\circ\) for the triangular cross-section.}
\label{tab:sameside}
\begingroup\setlength{\tabcolsep}{3pt}\renewcommand{\arraystretch}{1.15}
\begin{tabular*}{\linewidth}{@{\extracolsep{\fill}}llcccc@{}}
\toprule
\shortstack{Cross-\\section} & angles & regimes & \(R_i\) ratio & \(R_w\) ratio & \(\widehat{\mathcal L}\) ratio \\
Square & $46^\circ/48^\circ$ & E/E & 1.15--1.49 & 1.61--2.57 & 2.12--2.80 \\
Square & $48^\circ/52^\circ$ & E/E & 1.50--1.86 & 2.01--3.02 & 3.17--5.77 \\
Square & $42^\circ/44^\circ$ & N/N & 1.25--1.78 & 2.17--3.10 & 4.35--9.14 \\
Square & $44^\circ/46^\circ$ & N/E & 1.08--1.70 & 1.99--3.84 & 2.84--9.28 \\
Triangular & $62^\circ/66^\circ$ & E/E & 0.97--1.60 & 1.63--1.89 & 1.71--3.86 \\
Triangular & $58^\circ/62^\circ$ & N/E & 1.25--1.97 & 3.09--7.46 & 7.43--37.67 \\
\bottomrule
\end{tabular*}\endgroup
\end{minipage}
\end{table}

\clearpage
\section{Polynomial least-squares comparison}
\label{supp:polynomial}
For the square cross-section, the polynomial trial family is
\[
 u(x,y)=\sum_{\substack{0\le i,j\le k\\(i,j)\ne(0,0)}}c_{ij}T_i(x)T_j(y),
 \qquad k=8,12,16,20.
\]
Here \(k\) bounds the degree in each coordinate separately, so the trial space is spanned by \((k+1)^2-1\) tensor products of total degree up to \(2k\). Omitting \((i,j)=(0,0)\) sets the constant Chebyshev coefficient \(c_{00}\) to zero. Because the residuals involve only derivatives of \(u\), this removes the axial translation \(u\to u+C\) without a gauge penalty. It does not set \(u(0,0)=0\), since \(T_i(0)T_j(0)=\pm1\) for even \(i\) and \(j\) (for example \(T_2(0)=-1\)); plotted displacements and cap errors therefore subtract \(u(0,0)\) explicitly. Each angle and degree is an independent zero-initialized fit, with 80, 168, 288 or 440 coefficients. Fits cover contact angles \(30,38,40,42,44,46,48,52\) degrees for the square cross-section. They minimize the same residual objective as the networks, with \(w=10\), on 6000 interior and 2000 wall points drawn with seed 0 from the training laws in Sec.~\ref{app:floor}. Levenberg--Marquardt uses an analytic Jacobian, tolerances \(10^{-14}\), and at most 4000 function evaluations.

\begin{figure}[!htbp]
\centering
\includegraphics[width=\linewidth]{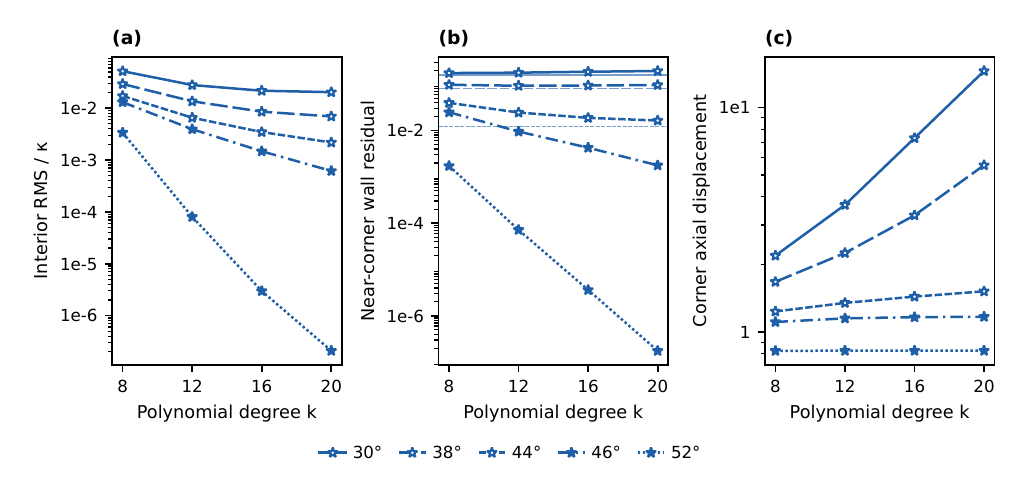}
\caption{Finite-degree dependence of saved polynomial least-squares fits in the square cross-section. Blue denotes the square cross-section, line patterns identify contact angle, and stars identify the polynomial trial family. Hollow symbols denote the nonexistence regime (\(\gamma<45^\circ\)); filled symbols denote the equilibrium regime (\(\gamma>45^\circ\)). (a) Relative interior RMS from Gauss order 256, (b) largest wall residual within \(10^{-3}\) of a corner, and (c) mean corner axial displacement, at degrees \(k=8,12,16,20\). Thin horizontal references in (b) are \(\delta=\cos\gamma-\cos\gamma_c\) in the nonexistence regime. Each angle and degree is an independent fit; connecting segments show the finite observations.}
\label{fig:polynomial_degree}
\end{figure}

Figure~\ref{fig:polynomial_degree} shows how the interior RMS, local wall maximum and corner axial displacement vary with degree. Table~\ref{tab:poly_floor} reports the degree-20 expected losses and their ratios to the certified bound. Quadrature uses coefficients from refits with the same settings, which reproduce the saved training losses to a relative \(2\times10^{-11}\). At \(44^\circ\), the closest angle, the ratio is 3.19, higher than at the other subcritical angles.

\clearpage
\begin{table}[!t]
\begin{minipage}{\linewidth}
\small
\caption{Expected loss \(\mathcal L\) of the degree-20 fits in the square cross-section and its ratio to the certified bound \(F\) under the evaluation law, at \(w=10\). Quadrature uses Gauss order 256 (Sec.~\ref{supp:quadrature}).}
\label{tab:poly_floor}
\begingroup\setlength{\tabcolsep}{3pt}\renewcommand{\arraystretch}{1.15}
\begin{tabular*}{\linewidth}{@{\extracolsep{\fill}}cccc@{}}
\toprule
\(\gamma\) & \(\mathcal L\) & \(F\) & \(\mathcal L/F\) \\
\midrule
$30^\circ$ & $3.93\times10^{-2}$ & $2.53\times10^{-2}$ & 1.55 \\
$38^\circ$ & $6.84\times10^{-3}$ & $4.50\times10^{-3}$ & 1.52 \\
$40^\circ$ & $3.04\times10^{-3}$ & $1.97\times10^{-3}$ & 1.54 \\
$42^\circ$ & $8.98\times10^{-4}$ & $5.44\times10^{-4}$ & 1.65 \\
$44^\circ$ & $9.94\times10^{-5}$ & $3.12\times10^{-5}$ & 3.19 \\
\bottomrule
\end{tabular*}\endgroup
\end{minipage}
\end{table}

\FloatBarrier
\section{Rounded-domain geometry and angle comparisons}
\label{supp:rounded}\label{app:rounded_geometry}\label{app:rounded_checks}
Let \(\Omega_\rho=[-1+\rho,1-\rho]^2+B_\rho\), where \(B_\rho\) is the radius-\(\rho\) disk and the plus sign denotes a Minkowski sum. The boundary is \(C^1\), with straight segments tangent to circular arcs. Its area, perimeter and prescribed curvature are
\[
 A_\rho=4-(4-\pi)\rho^2,\quad
 P_\rho=8-(8-2\pi)\rho,\quad
 \kappa=P_\rho\cos\gamma/A_\rho.
\]
For \(\rho\le\varepsilon\le1\), set \(D_\varepsilon=\Omega_\rho\cap\{x+y>2-\varepsilon\}\). Its wall \(\Gamma_\varepsilon\), cut \(\Sigma_\varepsilon\), and area are
\begin{align}
 |\Gamma_\varepsilon|&=2\varepsilon-(2-\pi/2)\rho,\\
 |\Sigma_\varepsilon|&=\sqrt2\,\varepsilon,\\
 |D_\varepsilon|&=\varepsilon^2/2-(1-\pi/4)\rho^2.
\end{align}
Four reflected copies have disjoint interiors and wetted walls, up to endpoints. On the top-right cut the outward normal is \((-1,-1)/\sqrt2\); on an arc it is the unit radial vector from its center.

All comparisons use \(\rho=0.005\) and \(w=10\). At \(40^\circ\), the cut \(\varepsilon=0.075\) gives
\[
 B=\cos\gamma|\Gamma|-|\Sigma|-\kappa|D|
 \simeq2.9005153\times10^{-3}>0.
\]
The positive deficit proves nonexistence by Finn's criterion \cite{supp-r7,supp-r20}. Under uniform probability measures the four-cut bound is
\begin{equation}\label{eq:roundedfloor}
 F(\varepsilon)=\frac{4[B(D_\varepsilon)]_+^2}
 {A_\rho|D_\varepsilon|+(P_\rho/w)|\Gamma_\varepsilon|}.
\end{equation}
Enumerating stationary points of the quotient and checking the endpoints gives \(\varepsilon=0.0612581625\) and \(F=2.84085435\times10^{-4}\). A bounded scalar search gives the same maximum. No positive deficit is found in this cut family at \(50^\circ\) or \(60^\circ\); existence at those comparison angles is not established here.

\begin{figure}[!htbp]
\centering
\includegraphics[width=\linewidth]{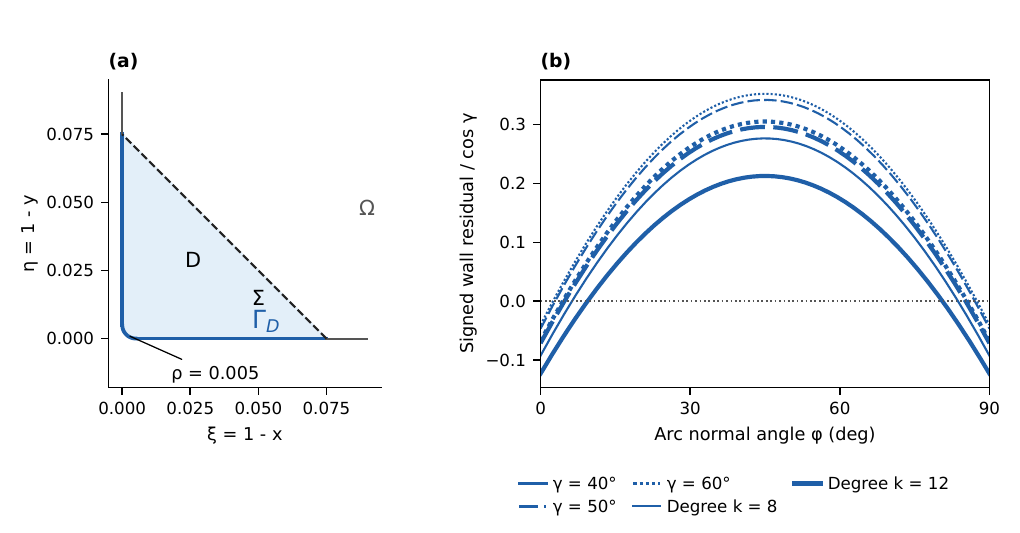}
\caption{Rounded-square cross-section and same-domain angle comparisons from Chebyshev polynomial least squares. (a) The top-right rounded corner at \(\rho=0.005\), in local coordinates \(\xi=1-x\), \(\eta=1-y\). Blue denotes the wetted boundary \(\Gamma_D\), the dashed segment is the artificial cut \(\Sigma\), and shading denotes \(D\). (b) Signed wall residual divided by \(\cos\gamma\) on the top-right circular arc, parameterized by outward-normal angle \(\varphi\). The six curves use the same domain, fitting and evaluation protocol: line pattern identifies \(\gamma=40^\circ,50^\circ,60^\circ\), and line thickness identifies polynomial degree \(k=8,12\). The \(40^\circ\) case has a proved nonexistence certificate; existence at the comparison angles \(50^\circ\) and \(60^\circ\) is not established here. Large arc residuals occur at all three angles.}
\label{fig:rounded}
\end{figure}

The polynomial trials have degrees 8 and 12, each initialized independently at zero. They minimize uniform-area mean \(f^2\) plus ten times uniform-arclength mean \(g^2\), with analytic derivatives and Jacobians. Levenberg--Marquardt allows 250 function evaluations and uses tolerances \(10^{-12}\); these fits terminate on relative cost reduction. The training rule for the rounded-square cross-section has 2752 interior and 288 wall nodes: tensor Gauss rules integrate a central rectangle, two side strips and four quarter disks, and separate rules integrate each straight wall and arc. The central, strip-radial, arc-angular, straight-wall and arc-wall orders are respectively \((32,12,20,48,24)\); the two evaluation sets use \((48,18,32,96,64)\) and \((64,24,48,128,128)\). In a side strip the long-direction order is the central order, and in a quarter disk the radial order is the strip order. Reported arc maxima use 513 nodes per quarter-circle; the plotted profiles use 1001. Table~\ref{tab:rounded} reports the six same-domain comparisons; Fig.~\ref{fig:rounded} shows the geometry and arc profiles.

\begin{table}[!htbp]
\begin{minipage}{\linewidth}
\small
\caption{Fits in the same rounded-square cross-section. All RMS values and expected losses use independent uniform-measure quadrature. Arc maxima are divided by \(\cos\gamma\). The last column is the unnormalized largest interior residual on a separate 48\,689-point probe, including points close to the arcs; it is a sampled maximum.}
\label{tab:rounded}

\setlength{\tabcolsep}{4pt}
\begingroup\setlength{\tabcolsep}{3pt}\renewcommand{\arraystretch}{1.15}
\begin{tabular*}{\linewidth}{@{\extracolsep{\fill}}ccccccc@{}}
\toprule
\(\gamma\) & \(k\) & \(\mathrm{RMS}(f)/\kappa\) & \(\mathrm{RMS}(g)/\cos\gamma\) & \(\mathcal L\) & \(\max_{\rm arc}|g|/\cos\gamma\) & \(\max_{\rm probe}|f|\) \\
$40^\circ$ & 8 & 0.0222 & 0.0224 & $4.1\times10^{-3}$ & 0.277 & 0.26 \\
$40^\circ$ & 12 & 0.0155 & 0.0183 & $2.52\times10^{-3}$ & 0.213 & 1.08 \\
$50^\circ$ & 8 & 0.0053 & 0.0166 & $1.19\times10^{-3}$ & 0.342 & 0.0622 \\
$50^\circ$ & 12 & 0.0060 & 0.0148 & $9.65\times10^{-4}$ & 0.296 & 0.59 \\
$60^\circ$ & 8 & 0.0045 & 0.0168 & $7.29\times10^{-4}$ & 0.352 & 0.122 \\
$60^\circ$ & 12 & 0.0047 & 0.0154 & $6.15\times10^{-4}$ & 0.305 & 0.277 \\
\bottomrule
\end{tabular*}\endgroup
\end{minipage}
\end{table}

The global RMS and localized peaks convey different information. At \(40^\circ\), the sampled interior peak grows from 0.260 to 1.08 as degree rises although its RMS decreases; the largest values lie close to the arc midpoint \(|x|=|y|\simeq0.99853553\). Large normalized wall peaks occur at all three angles. Arc concentration alone therefore does not identify nonexistence.

Local integration checks use the same \(\varepsilon=0.075\) cut at every angle and evaluate both sides of
\[
 \int_D f\,dA-\int_\Gamma g\,ds
 =\int_\Sigma Tu\cdot n_D\,ds+\cos\gamma|\Gamma|-\kappa|D|.
\]
This verifies the implemented flux balance; the nonexistence label comes from the analytic deficit. The optimized cutoff in Eq.~\ref{eq:roundedfloor} serves to report the certified lower bound, rather than change the common comparison region.

The unit disk at \(40^\circ\) provides a reference with a known solution, \(u=R-\sqrt{R^2-r^2}\), \(R=1/\cos\gamma\), and \(\kappa=2\cos\gamma\). Its degree-8/12 fits use 2560 interior and 96 wall training nodes, with radial/angular orders 32/20 in each quarter disk and 24 wall points per arc; evaluations use the two finer orders above, replacing the quarter-disk radial order by the central order. The relative cap errors are \(6.18\times10^{-4}\) and \(1.84\times10^{-5}\); expected losses are \(1.75\times10^{-5}\) and \(4.72\times10^{-8}\). This checks the polynomial residual implementation on a solvable domain.

\FloatBarrier
\section{Deterministic quadrature of expected losses}\label{supp:quadrature}

We compute the interior and wall mean-square residuals of the 141 networks and 32 polynomial fits (Sec.~\ref{supp:polynomial}) by deterministic Gauss--Legendre quadrature under the evaluation laws of Sec.~\ref{app:floor}. The uniform-area component uses the Duffy-collapsed Gauss rule of Sec.~\ref{supp:training} on the \(n\) center triangles. The vertex-cluster component uses the generating variables \(U\in[0,1]\) and \(\theta\in[-\alpha,\alpha]\). The wall rule uses \(v\in[0,1]\) on each half side, with \(z=\pm(1-v^2)\) and \(t=\pm(1-v^3)\), which keeps the integrand smooth up to the vertices. The probability weights sum to one. We use orders 32, 64 and 128 for convergence checks and order 256 for the fits in Table~1 and Table~\ref{tab:poly_floor}. Figure~3 uses order 128. The interior RMS in Figure~\ref{fig:polynomial_degree} uses order 256. The corner-weighted cap errors use the same area and vertex-cluster rules at order 128. Network derivatives up to second order are propagated analytically and agree with automatic differentiation to a relative \(2\times10^{-14}\). Polynomial derivatives use Chebyshev-series differentiation. The certified bounds were independently recomputed and agree with the cached values to a relative \(2\times10^{-11}\).

For the Table~1 losses the relative change from order 128 to 256 is at most \(6\times10^{-8}\). From order 64 to 128 it is at most \(3\times10^{-5}\) across the networks and \(4.2\times10^{-11}\) across the polynomials. As a cross-check, independent estimates on 40\,000 interior points (30\,000 uniform and 10\,000 corner-clustered) and 8000 wall points differ from the Table~1 quadrature losses by \(-1.5\%\) to \(+5.6\%\). The evaluation points use NumPy seed 12345. Table~\ref{tab:quadpop} gives the interior and wall terms for the Table~1 groups, and Table~\ref{tab:quadhead} lists the 15 fits behind the farther-subcritical rows. At \(w=10\) the wall term carries 79--99 percent of the loss. The six width-16 to 64 fits of the Width survey at \(40^\circ\), outside Table~1, give \(\mathcal L/F\) of 1.60--1.70.

\begin{table}[H]
\begin{minipage}{\linewidth}
\footnotesize
\caption{Gauss order 256 results for the fits in Table~1. The interior and wall terms \(\mathbb E_pf^2\) and \(w\,\mathbb E_qg^2\) sum to \(\mathcal L\), and \(F\) is the certified single-cut bound under the same evaluation law. Ranges are over fits.}
\label{tab:quadpop}
\begingroup\setlength{\tabcolsep}{2pt}\renewcommand{\arraystretch}{1.15}
\begin{tabular*}{\linewidth}{@{\extracolsep{\fill}}lcccccc@{}}
\toprule
\shortstack{Cross-\\section} & \(\gamma\) & \(w\) & Fits & \shortstack{Interior\\term} & \shortstack{Wall\\term} & \(\mathcal L/F\) \\
\midrule
Square & $30^\circ$ & 10 & 3 & $1.08\text{--}1.16\times10^{-3}$ & $3.94\text{--}4.12\times10^{-2}$ & 1.60--1.67 \\
Square & $38^\circ$ & 10 & 3 & $1.35\text{--}1.90\times10^{-4}$ & $6.93\text{--}7.29\times10^{-3}$ & 1.58--1.66 \\
Square & $40^\circ$ & 10 & 2 & $4.94\text{--}5.79\times10^{-5}$ & $3.05\text{--}3.27\times10^{-3}$ & 1.57--1.68 \\
Square & $42^\circ$ & 10 & 3 & $5.06\text{--}6.55\times10^{-5}$ & $8.93\text{--}9.43\times10^{-4}$ & 1.74--1.83 \\
Square & $44^\circ$ & 10 & 23 & $1.00\text{--}3.95\times10^{-5}$ & $0.78\text{--}1.84\times10^{-4}$ & 2.85--7.18 \\
Square & $44^\circ$ & 1 & 4 & $0.69\text{--}4.08\times10^{-5}$ & $0.20\text{--}1.05\times10^{-4}$ & 8.59--46.76 \\
Square & $44^\circ$ & 100 & 4 & $5.46\text{--}8.09\times10^{-5}$ & $6.35\text{--}6.55\times10^{-4}$ & 2.22--2.30 \\
Triangular & $45^\circ$ & 10 & 2 & $2.01\text{--}2.18\times10^{-3}$ & $7.01\text{--}7.02\times10^{-2}$ & 1.62--1.63 \\
Triangular & $54^\circ$ & 10 & 2 & $0.96\text{--}1.70\times10^{-4}$ & $7.96\text{--}8.27\times10^{-3}$ & 1.57--1.61 \\
Triangular & $58^\circ$ & 10 & 17 & $0.20\text{--}1.11\times10^{-4}$ & $4.95\text{--}9.62\times10^{-4}$ & 1.63--3.39 \\
\bottomrule
\end{tabular*}\endgroup
\end{minipage}
\end{table}

\begin{table}[H]
\begin{minipage}{\linewidth}
\footnotesize
\caption{The 15 fits of the farther-subcritical rows of Table~1, at \(w=10\). Gauss order 256 gives the interior and wall terms and their sum \(\mathcal L\).}
\label{tab:quadhead}
\begingroup\setlength{\tabcolsep}{2pt}\renewcommand{\arraystretch}{1.12}
\begin{tabular*}{\linewidth}{@{\extracolsep{\fill}}lccccccc@{}}
\toprule
\shortstack{Cross-\\section} & \(\gamma\) & Width & Seed & \(\mathcal L\) & \shortstack{Interior\\term} & \shortstack{Wall\\term} & \(\mathcal L/F\) \\
\midrule
Square & $30^\circ$ & 48 & 0 & $4.237\times10^{-2}$ & $1.15\times10^{-3}$ & $4.121\times10^{-2}$ & 1.671 \\
Square & $30^\circ$ & 48 & 1 & $4.047\times10^{-2}$ & $1.08\times10^{-3}$ & $3.940\times10^{-2}$ & 1.597 \\
Square & $30^\circ$ & 48 & 2 & $4.084\times10^{-2}$ & $1.16\times10^{-3}$ & $3.968\times10^{-2}$ & 1.611 \\
Square & $38^\circ$ & 48 & 0 & $7.451\times10^{-3}$ & $1.65\times10^{-4}$ & $7.286\times10^{-3}$ & 1.655 \\
Square & $38^\circ$ & 48 & 1 & $7.120\times10^{-3}$ & $1.90\times10^{-4}$ & $6.929\times10^{-3}$ & 1.581 \\
Square & $38^\circ$ & 48 & 2 & $7.162\times10^{-3}$ & $1.35\times10^{-4}$ & $7.026\times10^{-3}$ & 1.591 \\
Square & $40^\circ$ & 128 & 0 & $3.318\times10^{-3}$ & $4.94\times10^{-5}$ & $3.269\times10^{-3}$ & 1.680 \\
Square & $40^\circ$ & 128 & 1 & $3.109\times10^{-3}$ & $5.79\times10^{-5}$ & $3.051\times10^{-3}$ & 1.574 \\
Square & $42^\circ$ & 48 & 0 & $9.965\times10^{-4}$ & $5.34\times10^{-5}$ & $9.431\times10^{-4}$ & 1.833 \\
Square & $42^\circ$ & 48 & 1 & $9.939\times10^{-4}$ & $6.55\times10^{-5}$ & $9.284\times10^{-4}$ & 1.829 \\
Square & $42^\circ$ & 48 & 2 & $9.438\times10^{-4}$ & $5.06\times10^{-5}$ & $8.932\times10^{-4}$ & 1.736 \\
Triangular & $45^\circ$ & 48 & 0 & $7.217\times10^{-2}$ & $2.01\times10^{-3}$ & $7.016\times10^{-2}$ & 1.624 \\
Triangular & $45^\circ$ & 48 & 1 & $7.224\times10^{-2}$ & $2.18\times10^{-3}$ & $7.007\times10^{-2}$ & 1.626 \\
Triangular & $54^\circ$ & 48 & 0 & $8.364\times10^{-3}$ & $9.62\times10^{-5}$ & $8.268\times10^{-3}$ & 1.613 \\
Triangular & $54^\circ$ & 48 & 1 & $8.132\times10^{-3}$ & $1.70\times10^{-4}$ & $7.962\times10^{-3}$ & 1.569 \\
\bottomrule
\end{tabular*}\endgroup
\end{minipage}
\end{table}

\noindent\begin{minipage}{\linewidth}
\setlength{\parindent}{1.2em}\indent
Tables~\ref{tab:vertexslope} and~\ref{tab:polyslope} give the exact vertex flux \(|Tu|\) and slope \(|\nabla u|=|Tu|/(1-|Tu|^2)^{1/2}\) of the matched-capacity networks closest to the critical angle and of the polynomial fits below \(45^\circ\). All slopes are finite. For the polynomial fits the largest vertex slope grows with degree at each tested angle below \(45^\circ\), most strongly far from the critical angle. These are finite observations under a fixed budget; they do not establish a limit.

\begin{table}[H]
\begin{minipage}{\linewidth}
\footnotesize
\caption{Vertex flux and slope of the matched-capacity networks, seeds 3--7, near the critical angle. Ranges are over all vertices and fits of a row; the last column is the median over fits of the largest vertex slope.}
\label{tab:vertexslope}
\begingroup\setlength{\tabcolsep}{3pt}\renewcommand{\arraystretch}{1.12}
\begin{tabular*}{\linewidth}{@{\extracolsep{\fill}}llcccc@{}}
\toprule
Cross-section & Width & Fits & \(|Tu|\) at vertices & Vertex slope & \shortstack{Median of\\largest slope} \\
\midrule
Square, $44^\circ$ & 48 & 5 & 0.9958--0.9985 & 10.9--18.3 & 16.2 \\
Square, $44^\circ$ & 64 & 5 & 0.9967--0.9988 & 12.3--20.3 & 17.9 \\
Square, $44^\circ$ & 128 & 5 & 0.9972--0.9989 & 13.3--21.8 & 19.3 \\
Triangular, $58^\circ$ & 48 & 5 & 0.9985--0.9997 & 18.2--40.4 & 26.8 \\
Triangular, $58^\circ$ & 64 & 5 & 0.9971--0.9994 & 13.2--29.1 & 25.3 \\
Triangular, $58^\circ$ & 128 & 5 & 0.9990--0.9997 & 21.9--38.2 & 32.3 \\
\bottomrule
\end{tabular*}\endgroup
\end{minipage}
\end{table}
\end{minipage}

\begin{table}[H]
\begin{minipage}{\linewidth}
\footnotesize
\caption{Largest vertex slope \(|\nabla u|\) of the square-section polynomial fits of Sec.~\ref{supp:polynomial} by degree \(k\), and the corresponding largest vertex flux \(|Tu|\) at \(k=20\).}
\label{tab:polyslope}
\begingroup\setlength{\tabcolsep}{4pt}\renewcommand{\arraystretch}{1.12}
\begin{tabular*}{\linewidth}{@{\extracolsep{\fill}}cccccc@{}}
\toprule
\(\gamma\) & \(k=8\) & \(k=12\) & \(k=16\) & \(k=20\) & \(|Tu|\), \(k=20\) \\
\midrule
$30^\circ$ & 9.5 & 32.2 & 130.2 & 485.2 & 0.999998 \\
$38^\circ$ & 6.6 & 17.2 & 50.5 & 152.1 & 0.999978 \\
$40^\circ$ & 5.7 & 13.1 & 32.2 & 81.8 & 0.999925 \\
$42^\circ$ & 4.7 & 9.3 & 18.5 & 37.1 & 0.999638 \\
$44^\circ$ & 3.9 & 6.2 & 9.3 & 13.8 & 0.997403 \\
\bottomrule
\end{tabular*}\endgroup
\end{minipage}
\end{table}

\end{document}